\documentclass{PoS}
\usepackage{amsmath,amssymb, amsfonts}
\usepackage{subcaption}
\usepackage{caption}

\title{Towards a four-loop form factor}

\ShortTitle{Towards a four-loop form factor}

\author{\speaker{Rutger Boels},  Bernd A. Kniehl \thanks{supported by the German Science Foundation (DFG) within the Collaborative Research Center 676 ``Particles, Strings and the Early Universe"}\\
        II. Institut f\"ur Theoretische Physik, Universit\"at Hamburg\\ Luruper Chaussee 149, D- 22761 Hamburg, Germany\\
        E-mail: \email{Rutger.Boels@desy.de} and \email{Bernd.Kniehl@desy.de}}

\author{Gang Yang  \thanks{supported by a DFG grant in the framework of the SFB 647 ``Space-Time-Matter"}\\
	Institut f\"ur Physik, Humboldt Universit\"at zu Berlin \\ IRIS Geb\"aude, Zum Gro$\beta$en Windkanal 6, 12489 Berlin, Germany\\
	State Key Laboratory of Theoretical Physics, Institute of Theoretical Physics, \\Chinese Academy of Sciences, Beijing 100190, China\\
	E-mail: \email{Gang.Yang@physik.hu-berlin.de}}

\abstract{ The four-loop, two-point form factor contains the first non-planar correction to the lightlike cusp anomalous dimension. This anomalous dimension is a universal function which appears in many applications. Its planar part in ${\cal N}=4$ SYM is known, in principle, exactly from AdS/CFT and integrability while its non-planar part has been conjectured to vanish. The integrand of the form factor of the stress-tensor multiplet in ${\cal N}=4$ SYM including the non-planar part was obtained in previous work. 
We parametrise the difficulty of integrating this integrand. We have obtained a basis of master integrals for all integrals in the four-loop, two-point class in two ways. First, we computed an IBP reduction of the integrand of the ${\cal N}=4$ form factor using massive computer algebra ({\tt Reduze}). Second, we computed a list of master integrals based on methods of the {\tt Mint} package, suitably extended using {\tt Macaulay2} / {\tt Singular}. The master integrals obtained in both ways are consistent with some minor exceptions. The second method indicates that the master integrals apply beyond ${\cal N}=4$ SYM, in particular to QCD. The numerical integration of several of the master integrals will be reported and remaining obstacles will be outlined.}

\FullConference{12th International Symposium on Radiative Corrections (Radcor 2015) and LoopFest XIV (Radiative Corrections for the LHC and Future Colliders)\\
		15-19 June, 2015\\
		UCLA Department of Physics \& Astronomy Los Angeles, USA}

\begin{document}


\section{Introduction}
The computation of perturbative corrections is of vital importance to understand and explore the quantum nature of Nature. This holds for practical and theoretical as well as phenomenological and formal points of view. From a practical and phenomenological perspective, it is important to understand the structure of infrared (IR) divergences in intermediate results such as scattering amplitudes when computing corrections to Large Hadron Collider (LHC) observables. From a more formal theoretical perspective, one would like to obtain an understanding beyond perturbation theory. Consistency with known perturbative results in the latter focus are vital. In this proceedings contribution, we report on progress towards computing a quantity which appears in both of these developments: the lightlike cups anomalous dimension. 

The cusp anomalous dimension is a universal function which governs (leading) IR divergences in quantum field theory. It is a function of only the constants (coupling constants and group theory factors) in the theory: it has no kinematic dependence. Hence it is an ideal target for computation and analysis and a valuable testing ground for new computational technology. It has been calculated in Quantum Chromodynamics (QCD) to three-loop order \cite{Baikov:2009bg} (see also \cite{Gehrmann:2010ue}), and in the planar sector of $\mathcal{N}=4$ supersymmetric Yang-Mills (SYM) theory in principle to any loop order \cite{Beisert:2006ez}. In the latter case the developments are driven by a cluster of ideas known as the AdS/CFT correspondence \cite{Maldacena:1997re}. which is by far best understood in its planar limit. Both of these developments meet at four loops, where the first non-planar correction to the cusp anomalous dimension occurs. In fact, a conjecture exists \cite{Becher:2009qa} that this non-planar correction vanishes, see also \cite{Gardi:2009qi, Caron-Huot:2013fea}. The drive of the present work is to compute this quantity from perhaps the simplest observable that contains it: the Sudakov form factor with two on-shell legs in $\mathcal{N}=4$ SYM . 

Modern cutting-edge computation proceeds in several distinct phases. The first stage is integrand generation. This could be done by Feynman graph perturbation theory, but typically, even if the resulting expressions are manageable by computer algebra, the result is not in its simplest possible form. Especially for highly supersymmetric theories this is a problem as no off-shell, manifestly supersymmetric and Lorenz-invariant formulations of them exists. Instead, we have generated the integrand of the two-point four-loop form factor in previous work \cite{Boels:2012ew} using color-kinematic duality  \cite{Bern:2008qj, Bern:2010ue, Bern:2012uf} as an ansatz generator, up to a single remaining parameter which escaped all unitarity cuts used in that publication to pin down coefficients. Such obtained integrand takes a compact form and has only up to quadratic numerators. Typically, the generated integrand is expressed in terms of integrals for which modern integration methods are hard to apply directly. Moreover, there are typically very many integrals appearing. Therefore, this necessarily requires the second stage, where the integrand is reduced to a basis of master integrals by solving integration-by-parts (IBP) identities. Finally, in the third stage, the master integrals are integrated and the pieces are assembled into the full result. 

In this proceedings contribution, we report on the IBP reduction of the four-loop form factor in $\mathcal{N}=4$ SYM as well as on the basis of master integrals for this form factor in generic quantum field theories, such as QCD. This extends work at two \cite{vanNeerven:1985ja} and three loops \cite{Gehrmann:2011xn}.  For this, we used two different approaches. One is to solve the set of IBP relations for the  $\mathcal{N}=4$ SYM form factor explicitly through massive computer algebra. The other is to count the number of master integrals using techniques from computational algebraic geometry. The work reported in this talk has been published in \cite{Boels:2015yna}, to which the reader is also referred for more fine details of our results.

\section{IBP reduction}

An $L$-loop Feynman integral with $n$ ``indices'' $a_1, \ldots, a_n$ is an integral of the form
\begin{equation}\label{eq:deffeynmanint}
I(a_1, \ldots, a_n) \equiv \int d^D l_1 \ldots d^Dl_L  \left(1/D_1\right)^{a_1} \ldots  \left(1/D_n\right)^{a_n} ,
\end{equation}
where $D_i$ are inverse propagators. Integrals with the same positive indices are said to belong to the same sector. A fundamental property of Feynman integrals is that they obey IBP identities,
\begin{equation}
\int d^D l_1 \ldots d^Dl_L  \frac{\partial}{\partial l^\mu_i} \left( \textrm{integrand} \right) = 0 .
\end{equation}
By computing the derivative of the integrand, this equation translates into a relation between Feynman integrals with different indices. One can express integrals for a given integral topology in terms of the above set of Feynman integrals by constructing a complete set of propagators (quadratic expressions in momenta). Completeness means that any inner product of the momenta involved can be expressed as a linear combination of the chosen complete set of propagators. Once such a complete set is chosen, the IBP relations are a large set of linear relations on what amount to vectors given by ordered vectors of (in our case) integers, with rational coefficients.

The IBP relations allow one to reduce the set of integrals obtained by integrand generation to a simpler set, known as the `master' integrals \cite{Chetyrkin:1981qh, Tkachov:1981wb}. A key step of this so-called Laporta \cite{Laporta:2001dd} algorithm is to define a ranking of integrals by assigning them a score. Then the linear relations may be solved by, basically, Gaussian elimination, such that complicated integrals are expressed in terms of simpler ones. Various public and private implementations of Laporta's algorithm exist, such as {\tt AIR} \cite{Anastasiou:2004vj}, {\tt FIRE} \cite{Smirnov:2008iw, Smirnov:2013dia, Smirnov:2014hma} and {\tt Reduze} \cite{2010CoPhC.181.1293S, vonManteuffel:2012np}.  See also {\tt LiteRed} \cite{Lee:2012cn, Lee:2013mka} for an alternative approach to IBP reduction. We explored  {\tt FIRE}, {\tt Reduze} and {\tt LiteRed} in some detail for the four-loop form factor problem. Only {\tt Reduze} was able to solve for all integral topologies needed in this problem, after resolving a disk access pile-up issue in the currently public version of this code.

For the four-loop integrals under study, there are twelve propagators fixed by integral topology; six additional propagators must be added in order to find a complete set. One important technical result of our work  \cite{Boels:2015yna} is that the choice of the set of additional propagators can influence the performance of the IBP reduction algorithms dramatically. For instance, if the integral topology under study has graph symmetry, then an additional set of six propagators may be chosen such that the set reflects this symmetry manifestly. This allows for automatic simplifications during the reduction phase when using {\tt Reduze}. Typically, however, this leads to complicated expressions for these additional propagators. Alternatively, one can try to find a set of simple propagators, sacrificing manifest graph symmetry. These two choices behave differently under IBP reduction: while Reduze could solve the simple set of IBP relations, it got stuck in solving the symmetric set. The bottleneck is the size of messages being passed through the message-passing-interface protocol.  A prominent example where this occurs is integral topology $26$ from \cite{Boels:2012ew}. 

\subsection{Results}
We have obtained an explicit IBP reduction of all integrals appearing in the four-loop Sudakov form factor in \cite{Boels:2012ew}. The original integrals have up to quadratic numerators. After reduction, only one twelve-propagator, quadratic-numerator integral is left in topology $26$.\footnote{The topology numbers always refer to the graphics and tables in section 5 of \cite{Boels:2012ew}.} Single numerator integrals are more common, also with the full twelve propagators, see Table \ref{table:counting_Reduze}.  Although simpler, the basis of master integrals still contains integrals of considerable complexity. 

\begin{table}[ht]

\caption{Master integral statistics of obtained IBP reduction. $s$ represents the power of numerators.} 
\begin{subtable}{.5 \linewidth}
\caption{planar form factor} 
\centering
\begin{tabular}{c | c c c} 
\hline 
\# props & $s=0$ & s=1 &$ s=2$ \\ [0.5ex] 
\hline 
12 & 8 & 6 & 0 \\ 
11 & 18 & 2 & 0  \\
10 & 43 & 9 & 0 \\
9  & 49 & 1 & 0 \\
8  & 51 & 4  & 1  \\
7 & 25 & 0 & 0 \\
6 & 8 & 0 & 0 \\
5 & 0 & 0 & 0  \\ 
\hline 
sum & 203 & 22 & 1  \\ [1ex] 
\end{tabular}
\end{subtable}%
\begin{subtable}{.5\linewidth}
\caption{non-planar form factor} 
\centering
\begin{tabular}{c | c c c} 
\hline 
\# props & $s=0$ & s=1 &$ s=2$ \\ [0.5ex] 
\hline 
12 & 10 & 10 & 1 \\ 
11 & 13 & 3 & 0  \\
10 & 34 & 10 & 0 \\
9  & 29 & 1 & 0 \\
8  & 32 & 3  & 1  \\
7 & 13 & 0 & 0\\
6 & 7 & 0 & 0 \\
5 & 1 & 0 & 0  \\ 
\hline 
sum & 139 & 27 & 2  \\ [1ex] 
\end{tabular}

\end{subtable}
\label{table:counting_Reduze}
\end{table} 

A preliminary examination using mainly FIESTA \cite{Smirnov:2008py, Smirnov:2009pb, Smirnov:2013eza} based on sector decomposition \cite{Binoth:2000ps} as well as automated Mellin-Barnes integrals \cite{Czakon:2005rk,Gluza:2007rt} was able to compute almost all master integrals for the planar form factor---with the exception of three. These are integrals of topology $25$ with a linear numerator and without a linear numerator as well as the scalar integral of topology $30$ without numerator. The integrand for topology 30 reads
\begin{multline}
l_6^{-2} l_5^{-2} l_4^{-2} l_3^{-2} (-l_5 + p_1)^{-2} (-l_4 + l_5)^{-2} (-l_6 + p_2)^{-2} (l_3 - l_4)^{-2} (-l_4 + l_5 + l_6)^{-2}\\ \times(-l_3 + p_1 + p_2)^{-2} (-l_3 + l_4 - l_5 + p_1)^{-2} (-l_3 + l_4 - l_5 - l_6 + p_1 + p_2)^{-2} ,
\end{multline}
while that for topology 25 reads
\begin{multline}
l_6^{-2} l_5^{-2} l_4^{-2} l_3^{-2} (l_3 - l_4)^{-2} (l_5 + l_6)^{-2} (-l_6 + p_2)^{-2} (-l_4 + p_1)^{-2} (-l_3 + p_1 + p_2)^{-2} \\ \times(-l_4 - l_5 + p_1)^{-2} (-l_3 + l_4 + l_5 + p_2)^{-2} (-l_3 + l_4 - l_6 + p_2)^{-2} .
\end{multline}
Note that linear numerators inherently include a choice: different numerators may be related to the same master integral using the IBP reduction. 

We would like to mention that there are interesting cancellations after the IBP reduction. In particular, the remaining free parameter in the previously obtained integrand drops out of the reduced results, which shows that it is a truly free parameter.

\section{Masters from Mint}
Master integrals obtained by explicit IBP reduction within $\mathcal{N}=4$ SYM are at least a subset of the master integrals of the corresponding computation within QCD. For the four-loop form factor, we have made this more precise by studying a method for obtaining just the master integrals, without explicitly solving the IBP relations. This method was proposed in \cite{Lee:2013hzt}, building on earlier work in this direction in \cite{Baikov:2005nv}. The algorithm has also been incorporated in a public code, the {\tt Mint} package \cite{Lee:2013hzt}. We have applied this algorithm to the four-loop form factor integrals, swapping in different approaches to perform steps the {\tt Mint} package cannot perform in its current incarnation.


The basic idea is to count the number of master integrals by exploring only the analytic structure of the integral topology. Loosely speaking, for a given topology of $m$ propagators, the number of master integrals can be obtained by counting the number of proper critical points of the sum of first and second Symanzik polynomials
\begin{equation}
G(\vec\alpha) = U(\vec\alpha) + F(\vec\alpha) \,,
\end{equation} 
where the proper critical points are defined by
\begin{equation}
\label{eq:critical-point-condition}
\frac{\partial G}{\partial \alpha_i} = 0 \ \ \ (i = 1, \ldots, m) \qquad
\mbox{and} \qquad G \neq 0.
\end{equation}
The proper critical points can be found efficiently by computing the Gr\"obner basis of the corresponding ideal
\begin{equation}
\label{eq:ideal}
I = \left\langle \frac{\partial G}{\partial \alpha_1} \,, \, \ldots \,, \frac{\partial G}{\partial \alpha_m} \,, \, \alpha_0 G - 1 \right\rangle\, ,
\end{equation}
and then counting the number of irreducible monomials in the obtained Gr\"obner basis. 

This procedure has been implemented in the {\tt Mathematica} package {\tt Mint}  \cite{Lee:2013hzt}. It works smoothly for many simple examples, such as three-loop Sudakov form factors. However, in the four-loop case, two further problems emerge. First,  in many cases the computation of the Gr\"obner basis turns out to be too hard to do in {\tt Mathematica}, and we solve them by using Macaulay2 \cite{M2} and Singular \cite{DGPS}. The second problem is that, in a few cases, the critical points are non-isolated, in the sense that the set of critical points can form an affine variety of dimension $\geq 1$. Such cases cannot be handled  by {\tt Mint}, but can be solved with some extra work. We refer the reader to \cite{Boels:2015yna, Lee:2013hzt} for more details. 

Given the number of master integrals, one can then choose an explicit set of integrals, as long as they are independent of each other.

\subsection{Results}

\begin{table}[ht]
\caption{Master integral statistics of Mint basis.}
\centering
\begin{tabular}{c | c c c c c c c c} 
\hline 
\# props & 5 & 6 & 7 & 8 & 9 & 10 & 11 & 12 \\
\hline
\textrm{all simple } & 1 & 8 & 25 & 48 & 52 & 58 & 32 & 20 \\
\textrm{simple + one double} & 0 & 0 &  1 &   5 &  1  &  12 &  3  & 14 \\
\hline 
\end{tabular}
\label{table:counting_Mint}
\end{table} 

Combining all possible 34 topologies together, we obtain in total 280 master integrals for four-loop Sudakov form factors. They are classified in Table \ref{table:counting_Mint} according to the number of propagators and the power of propagators. Furthermore, 28 basis integrals are of propagator type and 116 basis integrals contain at least one sub-bubble topology. Only the remaining 136 basis integrals are `genuine' four-loop vertex integrals. The most challenging integrals are among the 34 basis integrals which contain 12 propagators. We have cross-checked with the reduction of {\tt Reduze} and find that all above 280 master integrals are independent. 

Since the method based on {\tt Mint} only relies on the topologies of the given integrals and applies to arbitrary numerators, the results are expected to apply to any theory, including QCD. We remind the reader that, in Table \ref{table:counting_Reduze}, the counting of {\tt Reduze} concerns only master integrals from the reduction of the ${\cal N}=4$ SYM form factor.

\subsubsection{An interesting mismatch between {\tt Mint} and {\tt Reduze}}

While the counting based on the {\tt Mint} method indeed provides a set of independent basis integrals, we find that the reduction of {\tt Reduze} tends to include more basis integrals. More concretely, we find nine corner integrals, containing only up to ten propagators, are taken as master integrals by {\tt Reduze} (as well as {\tt FIRE}), but are reducible according to {\tt Mint}. In particular, three of them are corner integrals of only eight propagators.

A few possibilities may explain this discrepancy.
First, there is no proof that IBP relations have included all possible integral relations, so new hidden relations might exist beyond those given by IBP. 
Second, the reduction setup of {\tt Reduze} and {\tt FIRE} requires a truncation of the numbers of propagators and numerators. Therefore, it may be possible that some further IBP relations are missing in the present setup.
Third, the implemented method based on {\tt Mint} may contain a possible loop hole, namely, it has not taken into account the possible critical points at infinity.\footnote{We would like to thank Roman Lee for pointing out this possibility to us.}
It would be very interesting to explore all these possibilities and understand the precise cause of the mismatch.

\section{Conclusions}
We have presented two important steps toward the integration of four-loop form factors. The first is to identify a basis of master integrals, valid for four-loop, two-point form factors in generic quantum field theories. This step consists of an approach through computational-algebraic-geometry methods to identify the master integrals. The second step is the explicit IBP reduction of the two-point, four-loop form factor in $\mathcal{N}=4$ SYM. This was done through massive computer algebra using a modified version of the {\tt Reduze} code. Of course, both steps can and have been compared and contrasted.

The next step will be the integration of the master integrals. This will allow one to compute both the planar and non-planar cusp anomalous dimensions at four loops. The planar case is a known result and will provide an important cross-check for the method. The non-planar case is presently unknown and would be highly desirable to obtain. Efforts in this direction are under way. 

A direction for which fundamental new technology seems to be required is to extend the explicit IBP reduction for the form factor to more general integral classes than those which appear in $\mathcal{N}=4$ SYM. QCD is an obvious goal here, but also reductions with doubled-up propagators would be very interesting: these appear for instance when applying dimensional recurrences \cite{Tarasov:1996br} or when exploring the method of quasi-finite basis integrals, see \cite{vonManteuffel:2014qoa} \cite{vonManteuffel:2015gxa} \cite{vonmanteuffelradcor}.

\acknowledgments
The authors would like to thank the organisers of the Radcor/Loopfest conference for the opportunity to present this material. This work was supported in part by the German Science Foundation (DFG) through the Collaborative Research Center SFB~676 ``Particles, Strings and the Early Universe: the Structure of Matter and Space-Time" and a DFG grant in the framework of the SFB~647 ``Space -- Time -- Matter".

\bibliographystyle{JHEP}

\bibliography{4LoopNPCuspNotes}



\end{document}